# *In Situ* Electron Energy-Loss Spectroscopy in Liquids

Megan E. Holtz,[1] Yingchao Yu,[2] Jie Gao,[2] Héctor D. Abruña,[2] David A. Muller[1,3]

[1] School of Applied and Engineering Physics, Cornell University, Ithaca, NY 14850

[2] Department of Chemistry and Chemical Biology, Cornell University, Ithaca, NY 14850

[3] Kavli Institute at Cornell for Nanoscale Science, Cornell University, Ithaca, NY 14850

Corresponding Author

Megan E. Holtz

School of Applied Physics, Cornell University

E13 Clark Hall

Ithaca, New York 14853-2501

Phone: (607) 255-0654

Fax: (607) 255-2643

Email: meh282@cornell.edu

**ABSTRACT**

*In situ* scanning transmission electron microscopy (STEM) through liquids is a promising approach for exploring biological and materials processes. However, options for *in situ* chemical identification are limited: X-ray analysis is precluded because the liquid cell holder shadows the detector, and electron energy-loss spectroscopy (EELS) is degraded by multiple scattering events in thick layers. Here, we explore the limits of EELS for studying chemical reactions in their native environments in real time and on the nanometer scale. The determination of the local electron density, optical gap and thickness of the liquid layer by valence EELS is demonstrated. By comparing theoretical and experimental plasmon energies, we find that liquids appear to follow the free-electron model that has been previously established for solids. Signals at energies below the optical gap and plasmon energy of the liquid provide a high signal-to-background ratio regime as demonstrated for $LiFePO_4$ in aqueous solution. The potential for using valence EELS to understand *in situ* STEM reactions is demonstrated for beam-induced deposition of metallic copper: as copper clusters grow, EELS develops low-loss peaks corresponding to metallic copper. From these techniques, *in situ* imaging and valence EELS offer insight into the local electronic structure of nanoparticles and chemical reactions.

**Introduction**

*In situ* imaging using microfluidic cells is a promising approach for exploring biological and materials processes that typically occur in liquid environments (Williamson, et al., 2003; de Jonge, et al., 2009; Zheng, et al., 2009; de Jonge & Ross, 2011). Flowing liquids through the transmission electron microscope (TEM) holder facilitates the introduction of new chemicals into the liquid cell, enabling the study of multi-stage reactions (Klein, et al., 2011a), such as nanoparticle growth (Zheng, et al., 2009; Tao & Salmeron, 2011; Evans, et al., 2011). Developments in holder circuitry also provide the groundwork for *in situ* electrochemical studies such as fuel cell and battery voltage cycling (Radisic, et al., 2006). Recently, graphene "aquariums" have enabled high-resolution imaging through a thin liquid layer, opening doorways to atomic resolution studies in liquid environments (Yuk, et al., 2012; Park, et al., 2012).

Despite the promise of the liquid cell approach, options for *in situ* chemical identification are limited. Energy dispersive X-ray (EDS) spectroscopy is precluded because the liquid cell holder shadows the detector: X-ray counts are dominated by the Ti of the tip of the holder. Core-loss electron energy-loss spectroscopy (EELS), which provides chemical information by probing inner-shell electron transitions, is quickly degraded in thick liquid layers due to multiple scattering events (Klein, et al., 2011b; Jungjohann, et al., 2012): lower-energy scattering events occur multiple times for individual electrons, obscuring the weaker higher-energy signal. While deconvolution algorithms can attempt to extract the singly scattered signal from a multiply scattered signal, deconvolution cannot recover the signal once the core-loss peak is overwhelmed by multiple lower-energy scattering events and Poisson noise from the background exceeds the counts in the edge of interest.

While it is expected that chemical identification from core-loss EELS will be quickly obscured by lower energy multiple scattering with increasing thickness, valence EELS, which offers electronic structure information, has little lower-energy scattering to obscure the signal. Here we explore the thickness limits of core-loss and valence EELS and demonstrate the capability of valence EELS to determine the local electron density, optical gap and thickness of the liquid for a variety of liquids such as those used in lithium-ion batteries (propylene carbonate), nanoparticle growth ($H_2O$ and $CuSO_4/H_2O$) and polymer precursors (ethylene glycol). Valence energy-filtered TEM (EFTEM) measurements can provide spatially resolved electronic structure information through liquids, and offer a high signal to background ratio for energies lower than the liquid plasmon peak. The potential for using valence EELS to understand *in situ* STEM reactions is established for beam-induced deposition of metallic copper.

**Materials and Methods**

A liquid-flow TEM holder incorporating a microfluidic cell was developed by Protochips, Inc. and is described by Klein *et al* (Klein, et al., 2011a). The holder houses a microfluidic cell consisting of two silicon-processed chips with thin (~50 nm) silicon nitride membranes that form a viewing window when the chips are overlapped. To ensure that the two 20-40 micron by 200 micron wide silicon nitride membranes overlap to form a viewing window with thin regions in every loading of the holder, the windows are oriented orthogonal to each other. The microfluidic cell is encapsulated in the titanium tip of the holder with o-rings to prevent any liquid from leaking into the vacuum. The liquid is able to flow in to and out of the holder and is fed by an external syringe pump. The liquid flow enables multi-stage reactions in the tip of the holder. Setup and holder schematics are given in the supplementary material (Fig. S1).

Imaging and EELS were performed using a monochromated FEI Tecnai F-20 STEM/TEM operated at 200 kV and equipped with a Gatan 865 HR-GIF spectrometer for EELS analysis.

While the monochromator was not employed here, the study did benefit from the system's improved energy stability. EELS was performed, both in the STEM geometry, compatible with high-angle annular dark field (HAADF) imaging, and in EFTEM mode. The convergence and collection angles for STEM EELS were 9-10 mrad and 20-30 mrad, respectively. HAADF-STEM was used for its high signal-to-background ratio for particles with high-Z offered by the dark field detector, and its insensitivity to chromatic blur from energy losses in the sample. EFTEM with a 5 eV slit was explored as an option for short-time acquisitions of large images in relatively thin liquid layers. Since the energy slit was narrow, chromatic blur did not limit our spatial resolution, and we did not find it necessary to use an objective aperture in EFTEM mode.

**Results and Discussion**

Spatial resolution and contrast are dependent on the thickness of liquid layers (Reimer & Kohl, 2008). Due to multiple scattering, the electron beam broadens as it travels through a thick liquid layer (Demers, et al., 2012). There is a top-bottom effect where the resolution of particles imaged by STEM on the bottom of a thick liquid layer is severely degraded due to beam broadening (Reimer & Kohl, 2008). On the other hand, if the STEM probe interacts with the particles before the liquid layer, it will merely scatter some of the beam off of the HAADF detector and thus lower the intensity and the contrast of the signal, but not its resolution. To account for the top-bottom effect, STEM imaging of nanoparticles was performed on the entrance nitride membrane of the microfluidic cell to minimize probe-broadening effects. For energy-filtered TEM imaging, the opposite is true, and the best images are obtained for particles on the exit surface.

*Thickness Determination by EELS*

Investigating the thickness of the liquid layer is critical for understanding what is resolvable through liquids. EELS offers a convenient way to estimate the thickness of the layer traversed (*t*) in terms of the mean free path (*λ*) using Beer's law (Eq. 1), which predicts an exponential decay of the unscattered incident beam (Egerton, 2011):

$$I = I_0 \exp(-t/\lambda) \qquad \text{Eq. 1}$$

The thickness, in terms of the inelastic mean free path ($t/\lambda$), is measured by determining the ratio of the number of unscattered electrons (*I*) in the zero-loss peak (ZLP) to the total number of incident electrons ($I_0$). A large dispersion (0.3 eV) and large collection angle (>20 mrad) are used to detect most of the scattered electrons. However, if all the scattered electrons over the energy range acquired are not collected, a power law fit to the tail of the spectrum can be used to extrapolate the curve for a more accurate estimate of $I_o$. For EELS acquired out to 510 eV, the power law extrapolation introduces a 1% correction above $t/\lambda = 7$ in the liquid and grows to 4% for $t/\lambda = 9$. All thickness values reported here were determined using Beer's Law after extending the spectrum until it approached zero using a power law fit to the tail.

The conversion from thickness in terms of the inelastic mean free path (*λ*) to the thickness (t) in nm is done by estimating *λ* from the plasmon scattering. Here we estimate *λ* assuming that plasmon losses are the dominant loss mechanism in the sample, which, from simple inspection of a single scattering measurement, is a good approximation. The simple single-pole plasmon model, which is equivalent to assuming that the electrons in the fluid behave as free electrons, turns out to be reasonably valid as discussed later, and gives the simple result (Egerton, 2011):

$$\lambda = \frac{a_0}{\gamma \theta_{Ep} \ln(\theta_c/\theta_{Ep})} \qquad \text{Eq. 2}$$

where $a_0$ is the Bohr radius, $\gamma$ is the Lorentz factor, $\theta_c$ is the collection angle (~20 mrad). In addition, we define

$$\theta_{Ep} = \frac{E_p}{\gamma m_e v^2} = \frac{E_p}{E_i(1+\gamma^{-1})} \qquad \text{Eq. 3}$$

where $m_e$ and $v$ are the rest mass and velocity of the electron, respectively. $E_i$ is the incident energy (200 keV) and $E_p$ is the plasmon energy. The derivation of this formula assumes the Born approximation – the incident field is the primary field causing electron oscillation – and is thus only valid over the range where $E_i \gg E_p$. In general, this is true for a 200 keV incident beam and assuming that, on average, the inelastic scattering events lose the plasmon energy. This method gives a weighted average of $\lambda$ for the liquid and the nitride membranes. However, the nitride membranes are thin compared to the liquid layer. While Eq. 2 can provide reasonable values of $\lambda$ for free-electron materials, it tends to overestimate $\lambda$ by a few percent (Egerton, 2011), but it is preferred since it avoids requiring *a priori* knowledge of material parameters such as effective atomic number and relies only upon the plasmon energy, incident beam energy, and collection angle, which are all experimentally measurable and not free fitting parameters. For collection angles greater than 20 mrad, the dependency of $\lambda$ on the collection angle flattens off (Malis, et al., 1988; Botton, et al., 1995; Egerton, 2011; Zhang, et al., 2012), so a large collection angle prevents small experimental differences from drastically changing results, and allows the collection of most of the scattered electrons. More elaborate empirical fits have been attempted (Iakoubovskii, et al., 2008; Zhang, et al., 2012), but for the thicknesses typical in our liquid cell and collection angles larger than 20 mrad, the differences are less than 10%. As the characteristic elastic scattering angles are much larger than the characteristic inelastic scattering angles, the convolution of multiple elastic and inelastic scattering in the liquid will result in a measurement that again satisfies Beer's law for a fully-angle-integrated inelastic distribution (see the discussion leading to equation (3.112) in Egerton (Egerton, 2011)). Thus, from the EELS analysis, the total thickness of the liquid cell can be estimated from equation 1.

From STEM-EELS measurements, the calculated thicknesses for various locations across the viewing membrane are plotted as points in Fig. 1(a-b) with the contours representing a parabolic fit. In Fig. 1(a), 25 micron wide viewing windows with a water liquid layer are shown with a spacer ~250 nm thick, and in Fig. 1(b), 50 micron wide viewing windows with ethylene glycol are shown with a spacer ~500 nm thick. As expected, the increase in thickness seen in the center of the viewing membrane is caused by the $SiN_x$ membranes bowing out due to the pressure differential between the cell and TEM vacuum. A schematic of the bowing of the nitride membranes is shown in Fig. 1(c), representing two independent orthogonal parabolas. The total thickness is found from the distance between the top and bottom membranes, as shown in Fig. 1(d). The thickness profile in Fig. 1(d) corresponds to the fitted contour lines shown in Fig. 1(b). We find that the thickness is smallest at the corners, where the limiting factor is the spacer. However, in the center of the window, both the spacer and the bulging of the membranes contribute to an increased thickness, with thicker layers (~1 μm) obtained with the wider 40 μm windows and with thinner layers (~650 nm) for the narrower 25 μm windows. Thus, the thickness values appear reasonable from the geometry and are consistent with previously reported thicknesses (Klein, et al., 2011a; Jungjohann, et al., 2012). From this, it is clear that in order to achieve the highest resolution, one needs not only a thin spacing between the two chips, but also a narrow viewing window to limit bulging.

*Core-Loss EELS*

Core-loss EELS through various thicknesses of water and ethylene glycol from different locations across the window are shown in Fig. 2(a-b). The spectra are taken with a large energy dispersion of 0.3 eV, enabling the simultaneous collection of the core-loss signal along with the low-loss, providing a thickness and core-loss measurement from the identical region under identical conditions. As the liquid becomes thicker, increased multiple inelastic scattering events occur through the layer, degrading the core-loss signal in two ways. Firstly, the signal from the near-edge region of the core-loss peak will be multiply scattered to higher energies, reducing the jump ratio, and secondly, repeated losses to valence excitations will add multiple bulk plasmon peaks to higher energies, overwhelming the core-loss signal. With increasing liquid layer thickness, chemical identification provided by core-loss EELS in the weak, high-energy signal is quickly obscured by lower-energy plural scattering events. Thus, we find that core-loss EELS is only possible in liquid layers that are thinner than two to three times the inelastic mean free path, $\lambda$. Here the oxygen-K edge is only resolvable in the thinnest water layer ($t/\lambda \sim 2.7$) shown in Fig. 2(a).

Fourier-Log deconvolution, a commonly used method to remove the multiple scattering signal from EELS (Johnson & Spence, 1974; Spence, 1979; Egerton, 2011), helps define core-loss peaks from the background in moderately thin samples ($t/\lambda < 3$) as shown in Fig. S2. For reliable deconvolution of the spectra, the collection angles are bigger than the convergence angle and the largest plasmon single scattering angle, which renders the effect of subsequent elastic scattering on the inelastic ratios negligible. However, once the peak is obscured by lower-energy multiple scattering events, deconvolution cannot extract the signal and, in addition, noise amplified across large energy ranges (ringing artifacts) makes the spectrum uninterpretable.

*Valence EELS*

Valence EELS, which provides information about the local electronic structure of the material through collective excitations and interband transitions, represent the lowest-energy excitations observable by current instruments. They are also well separated from the lower-lying phonon modes, and thus are far less affected by plural scattering from lower energy transitions than core-loss excitations. Experimentally, we observe that the valence EELS signal, such as the optical gap, is resolvable up to $t/\lambda \sim 6\text{-}7$ in the liquid as seen in the insets of Fig. 2(a-b). It should be mentioned that valence EELS is resolvable in liquid layers roughly 2-3 times thicker than is possible for core-loss EELS. The limiting liquid thickness in which valence EELS is resolvable is predicted by its Poisson distribution. For valence EELS, the probability $P$ of a single scattering event occurring over the interval of the thickness of the liquid is given by the first order Poisson distribution (Fig. 2(c)):

$$P = (t/\lambda)\exp(-t/\lambda) \qquad \text{Eq. 4}$$

which has a maximum at $t/\lambda = 1$ where the probability of scattering once is 36.8%. Since the directly interpretable features of EELS come from the singly-scattered signal, we expect to be able to resolve features when the single scattering signal is at least 1% of the total signal, which occurs in liquid layers with thicknesses over the range of $t/\lambda = 0.01$ to 6.5 as shown in Fig. 2(c). This model provides plausibility that EELS can provide valence information in fairly thick TEM specimens (up to ~660 nm of water), even when core-loss information is lost.

We have performed valence EELS measurements on a variety of liquids likely to be used in *in situ* experiments, as shown in Fig. 3(a). The determination of the optical gap, plasmon energy and local electron density of the liquid by valence EELS is demonstrated for propylene carbonate (a commonly used solvent in lithium ion battery applications), ethylene glycol (a polymer precursor), water, and copper sulfate $CuSO_4/H_2O$ (studied for applications in nanoparticle growth and electroplating). As a control experiment, EELS from a 50 nm thick $SiN_x$ membrane is also shown in Fig. 3. Plotted on the same scale, it is clear that the liquid in the $SiN_x$/liquid/$SiN_x$ structure dominates the EELS signal.

A summary of the properties of the liquids is shown in Table 1. The mean free path is calculated from Eq. 2 as discussed earlier. The local electron density $n_o$ (valence electrons/m$^3$) is found using the free-electron model for the plasmon energy,

$$E_p = \hbar \sqrt{\frac{n_o e^2}{\varepsilon_0 m_e}} \quad \text{Eq. 5}$$

where $\hbar$ is Planck's constant divided by $2\pi$, $e$ is the elementary charge, $\varepsilon_0$ is the permittivity of free space, and $m_e$ is the mass of the electron. Alternatively, the theoretical plasmon energy can be calculated from bulk materials properties by estimating the local electron density as $n_o = n_e N_A \rho / A$ where $n_e$ is the number of valence electrons per molecule, $N_A$ is Avogadro's number, $\rho$ is the density of the material, and $A$ is the atomic weight. The theoretical versus experimental plasmon energies are shown in Fig. 3(b), compared to solids listed in Egerton (Egerton, 1986). For the solids, a linear fit with a slope of 1 is found, and the lines corresponding to one standard error margins are shown. The theoretical plasmon energy of the liquids matches reasonably well with the plasmon energy measured by EELS, confirming that the liquid behaves similarly to solids under the free electron model. Optical measurements have corroborated that the 21 eV peak in water is a collective electron oscillation (Heller, et al., 1974).

While it might seem surprising that a liquid responds to an external charge disturbance in a manner closer to that of a free-electron metal than an isolated molecule, it should not be unexpected considering that most insulators behave similarly as well (Egerton, 2011). The key ingredients for collective oscillations are effective screening and a uniform background over the appropriate length scales. For instance, in water the Thomas-Fermi screening length (Ashcroft & Mermin, 1976) for the valence electrons is ~0.4A, smaller than the size of the molecule, which has 10 electrons/molecule. Because there can be screening of electrons even on an individual water molecule, the first condition for collective oscillations is met. However, an isolated water molecule will not appear as a uniform density of charge at long wavelengths, while a liquid will – this second condition is also important and is discussed in the next paragraph. A plasmon is a charge density fluctuation made of up virtual excitations of electron hole pairs in a collective manner leading to a net displacement at a given wavelength of the entire Fermi Sphere (the full momentum-dependent electronic structure of occupied states), so the energy for the electron-hole pair excitations must be less than the plasmon energy, $E_p$, i.e. in an insulator with an optical gap $E_g$, $E_g<<E_p$. This is a necessary condition for electrons in insulators (and liquids) to behave like free electrons: single particle interactions with energies around the optical gap will then not significantly displace the measured plasmon energy from the free plasmon energy (Egerton, 2011). As the energy of the virtual excitations increases, their lifetime must decrease, which from the energy-time uncertainty principle leads to a broadening of the resulting plasmon peak. Additional damping can occur depending on the exact details of the single particle electronic structure at those energies. Metals and small-gap semiconductors would then be expected to have sharper plasmon peaks than large gap insulators. The plasmon line width in water (10 eV) is not that different from silica, silicon nitride or most other insulators.

The second condition for collective oscillations is that the density sampled is uniform. This turns out to be met for liquids, and can be understood by considering the lifetime of the plasmons, which have energy spreads of around 10 eV. From the energy-time uncertainty principle, the spread in energy corresponds to a lifetime of $7\times10^{-17}$ s. As an order of magnitude estimate, the charge oscillation spreads out at the speed of sound in the electron gas, which has an upper bound given by the Fermi velocity. Estimating this from the measured valence band width of ~ 5 eV, the plasmon travels ~ 0.9 Å, which is less than the scale of a water molecule, and provides evidence that the sampled electron density is uniform. The plasmon travel distance and lifetime correspond well to results calculated from time-dependent density functional theory, which shows ~1 Å distances and ~7-10 × $10^{-17}$ s lifetimes (Tavernelli, 2006). Although the oscillation does not propagate beyond a single molecule, collective motion is still possible because the Thomas-Fermi screening length of the electrons in the molecule (~0.4 Å) is smaller than the size of the molecule, allowing for electron interaction and screening on the same molecule (as there is more than one valence electron per molecule). Furthermore, the oscillating water molecule is surrounded by other water molecules of equal density and polarizability, so a relatively uniform electron density is sampled even at longer length scales, which satisfies the compressibility sum rule (Pines & Nozie`res, 1989), forcing the frequency of the collective displacement of the Fermi sphere to the free-electron limit at small momentum transfer. The details of the single particle excitations then determine the lifetimes of the quasiparticle excitations, which are very short lived in molecules compared to simple metals with more traditional single-particle band structures. The free electron formula underlying Eq. 2 and Eq. 5 will hold so long as the collective excitation cross sections exceed the single particle transitions and the valence orbitals are relatively uniform over the volume sampled by the excited state. The approximation should work well for *s* and *p* orbitals, but less well for the localized, narrow *d* and *f* states.

*Measurement of the Optical Gap*

The optical gap of a fluid is the minimum amount of energy required to excite an electron from the highest occupied molecular orbital to the lowest unoccupied state in the liquid, usually an excitonic state. Many liquids have relatively large optical gaps, while chemically active nanoparticles tend to exhibit strong excitations at lower energies. Thus the energy loss region below the optical gap can act as a transparent window through the liquid, where nanoparticle excitations can be observed with a relatively low background from the thick liquid layer, which will be discussed later for $LiFePO_4$. Furthermore, when the optical gap of the liquid is of interest, valence EELS enables a reliable determination of the optical gap of extremely small and/or localized amounts of liquid (~zeptoliters, zL; $10^{-21}$ L). The optical analog of EELS, ultra-violet and visible absorption spectroscopy (UV-VIS), can distinguish the optical gaps of liquids but only in bulk quantities (~mL). This demonstrates the strength of using valence EELS to qualitatively analyze the various types of liquids even for extremely small, localized volumes (~zL), although either probe broadening (Demers, et al., 2012) or delocalization (Muller & Silcox, 1995) will limit spatial resolution to the nm regime. This may provide details on how the electronic structure of a system changes during a chemical reaction if the species are changing across the viewing window, as we will demonstrate later with electron beam-induced reduction of copper sulfate to copper.

Local measurements of the optical gap of the liquid are possible using EELS, provided that the optical gap is larger than 3-5 times the width of the ZLP in order to avoid overlap and errors in background extrapolation. To find the location of the optical gap, we linearly fit the flat region after the ZLP and the sloped region after the optical gap and find the point of intersection between the two regions. For systems where the ZLP is not cleanly resolved from the optical excitations, more elaborate fitting functions are needed. In the copper sulfate solution, Cherenkov radiation is present, resulting in an increase in the intensity below the optical gap due to the electrons travelling faster than the speed of light in the medium (as discussed later). Because the raised background level would give an overestimate of the optical gap, we instead fit to the pre-zero-loss dark level and the sloped region after the optical gap and find that intersection. The precision of the optical gap measurement is limited by the energy resolution determined by the full width at half maximum of the ZLP as the spectrum is convolved with the ZLP; this shifts the onset by a value on the order of half the ZLP width. The optical gap measurements agree with UV-VIS measurements of pure liquids, but for the 10 mM copper sulfate solution the optical gap measured by EELS is between copper sulfate and water, likely because the scattering probability from the low concentration of $CuSO_4$ is much lower in the thin liquid layer.

There are two energy regimes where it is possible to maximize the signal to background ratio for an EELS measurement of a nanoparticle in a liquid. One regime is well past the bulk plasmon, where the core-loss signal, as discussed earlier, is only resolvable in liquid layers less than 3 $t/\lambda$ thick. The other regime is below the first plasmon of the liquid and, particularly, in the optical gap of the liquid. In the optical gap, there is no energy loss mechanism and hence little scattering from the liquid at those energy losses, thus offering an energy window with little background inelastic scattering. Here we explore the former regime by core-loss STEM-EELS and the latter regime by energy-filtered TEM (EFTEM) of $LiFePO_4$ nanoparticles in a $t/\lambda = 1.7$ (180 nm) thick layer of aqueous solution, as shown in Fig. 4. Core-loss EELS of the nanoparticles, compared to the background aqueous solution, are shown in Fig. 4(a), where the iron peak appears for the nanoparticles while the oxygen peak is observed for both the liquid and the nanoparticles. The EFTEM images for 0, 5, 10, and 25 eV are shown with a 5 eV energy selecting slit. At 0 eV and 25 eV in Fig. 4(b and c), the signal is dominated by the liquid and the nanoparticles show little interesting information of their own. However, in the optical gap of the liquid at 5 eV in Fig. 4(d), some of the nanoparticles have energy transitions corresponding to Li-poor regions (Sigle, et al., 2009; Moreau, et al., 2009; Kinyanjui, et al., 2010; Brunetti, et al., 2011). At 10 eV in Fig. 4(e), where there is still low liquid background, different regions of the nanoparticles exhibit transitions. This is highlighted in Fig. 4(f) where green and red correspond to the 5 eV and 10 eV signals. From this, we can quickly diagnose that the nanoparticles are inhomogeneous, and we can tell that the $FePO_4$ is generating the 5 eV signal. Conversely, when there is more 10 eV signal than 5 eV signal (as seen colored red) the particles are Li-rich.

In Situ *Deposition of Copper*

The potential for EELS in understanding *in situ* chemical reactions is demonstrated by the electron beam-induced deposition of metallic clusters of copper on the top $SiN_x$ membrane window during imaging. A 10 mM $CuSO_4/H_2O$ solution was irradiated at a beam dose rate of 30 e-/nm$^2$s. As the electrons impinge on the sample, the copper cations are reduced and clusters form and grow, as shown in Fig. 5(a). Valence EELS, in Fig. 5(b), confirms the growth of metallic copper: in the regions of zero to low deposition, the plasmon feature looks much like that of water. On the other hand, in regions of heavy copper deposition, peaks corresponding to metallic copper appear around 17 eV. For this study, the EELS delocalization through the thick liquid is less than the resolution due to beam spreading.

Additionally, the $CuSO_4$ solution can be distinguished from water by the two-fold increase in intensity seen below the optical gap. From the optical absorption data, there is a 1.5 eV peak in the CuSO$_4$ solution, corresponding to a Cu$^{2+}$ e$_g$ to t$_{2g}$ transition, which gives it a blue color (Jancso, 2005). With an energy resolution of 0.6 eV, the tail of the ZLP overwhelms this peak and it does not appear in the spectra. The detection of this peak may be possible with a monochromated beam, which would illustrate the advantages of high-energy resolution for the identification of a wider range of optical fingerprints. Although metallic copper is expected to produce a signal in the optical gap of the liquid, the signal in this region is identical in both the solution with and without metallic copper, so it appears to be uncorrelated with the presence of metallic copper. This is understandable considering that the amount of metallic copper is actually small – by comparing the area under the plasmon peaks of the low copper deposition to heavy copper deposition, the contribution from metallic copper is roughly 20% of the total signal.

Due to the flat nature of the below-gap intensity, and because it is uncorrelated with metallic copper, we believe it corresponds to Cherenkov radiation. Cherenkov radiation occurs when the charged electrons pass through a medium faster than the speed of light in the medium, $v > c/n$, where $v$ is the relativistic speed of the particle, $c$ is the speed of light in vacuum and $n = \sqrt{\varepsilon}$ where $n$ is the index of refraction and $\varepsilon$ is the frequency dependent dielectric constant in the medium. Thus, for 200 keV electrons, we expect Cherenkov radiation in materials with index of refraction $n$ greater than 1.44 (or for 300 keV electrons, $n > 1.29$). The appearance of Cherenkov radiation in the CuSO$_4$ solution indicates the increased index of refraction of CuSO$_4$ above 1.44 compared to pure water with index of refraction at 1.33. The small amount of Cherenkov radiation seen in water may be due to impurities. This finding follows the Frank-Tamm result, which states the energy lost per unit path of the particles is proportional to $[1-(c^2/\varepsilon v^2)]$ (Jackson, 1999). This background is likely to be more pronounced at 300 keV, where even pure water will display a Cherenkov background.

**Conclusion**

Due to multiple scattering, spatial resolution is dependent on the liquid layer thickness. Here we determine the thickness regimes where core-loss EELS and valence EELS are possible and useful. Chemical identification by core-loss EELS is only practical through < 3 inelastic mean free paths (~300 nm at 200 keV), but valence EELS still provides information to much larger thicknesses, < 6-7 inelastic mean free paths (~600-700 nm at 200 keV). Valence EELS gives insight into electronic structure properties such as the local electronic density and optical gap, and we find that plasmon scattering in the liquid follows the free electron model. Looking for low-energy signals below the plasmon energy and especially in the optical gap of the liquid provides a high signal to background ratio, which is demonstrated for valence EFTEM of $LiFePO_4$. Valence EELS through liquids is demonstrated to be a useful tool for quantifying liquid layers and changes in liquids during *in situ* chemical reactions. For electron beam-induced deposition of copper, plasmon peaks indicate the presence of metallic copper in heavy growth regions and Cherenkov radiation below the optical gap identifies the higher dielectric constant of the copper sulfate solution compared to water. EELS, and especially valence EELS, is well suited for providing chemical information about the products and reactants of chemical reactions studied by *in situ* TEM.

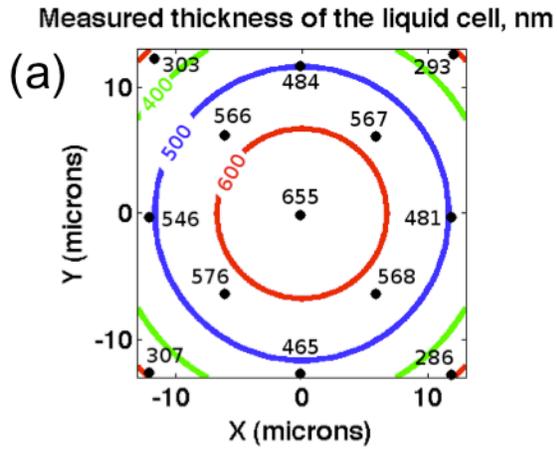
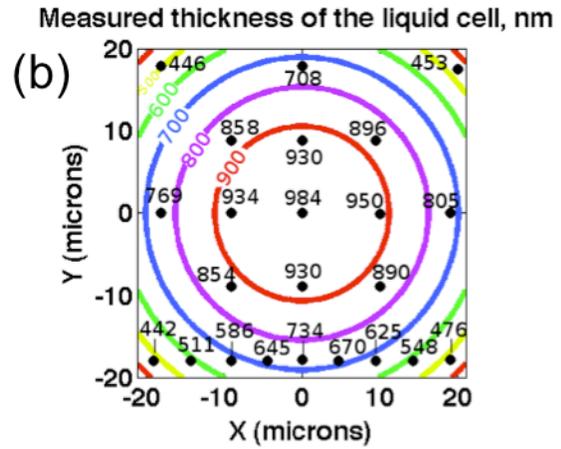
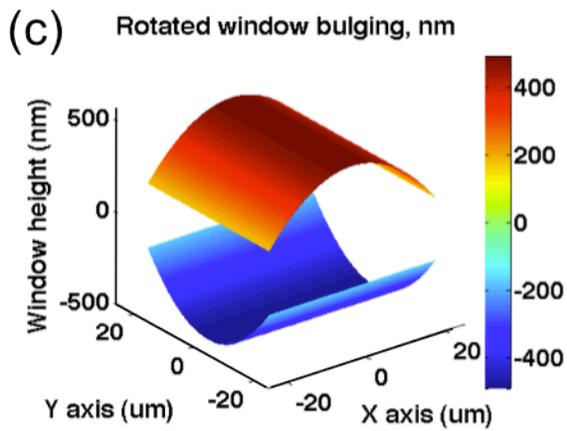
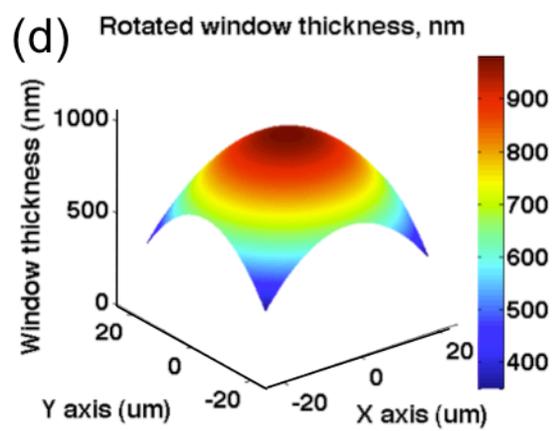

**Figure 1.** Measurement of the variations in liquid thickness due to the bulging of the windows. STEM-EELS measurements (points) with a parabolic fit (contours) are shown for (a) water between 25 micron wide windows with a 150 nm Au spacer and (b) ethylene glycol between 40 um wide windows with a 500 nm SU-8 spacer. (c) A 3-dimensional representation of the bulging of the two rotated $SiN_x$ membranes, assuming each window bows out as an independent orthogonal parabola and (d) the total thickness between the windows from (c). The profile shown in (d) corresponds to the contour lines in (b).

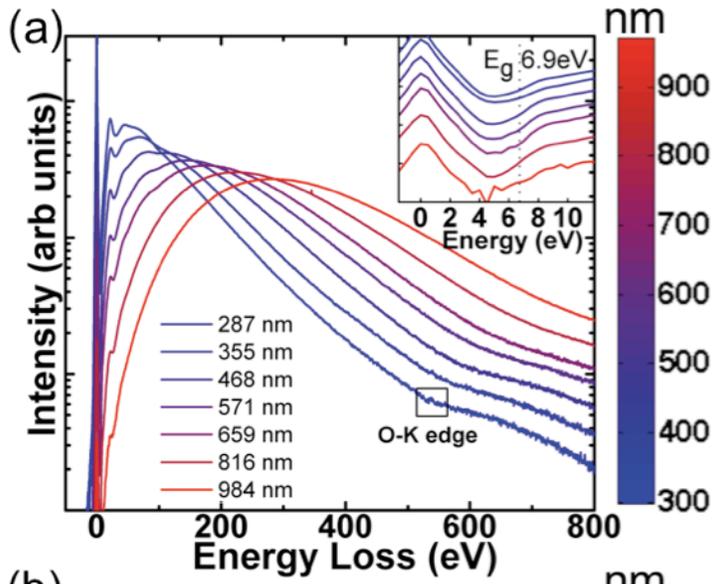

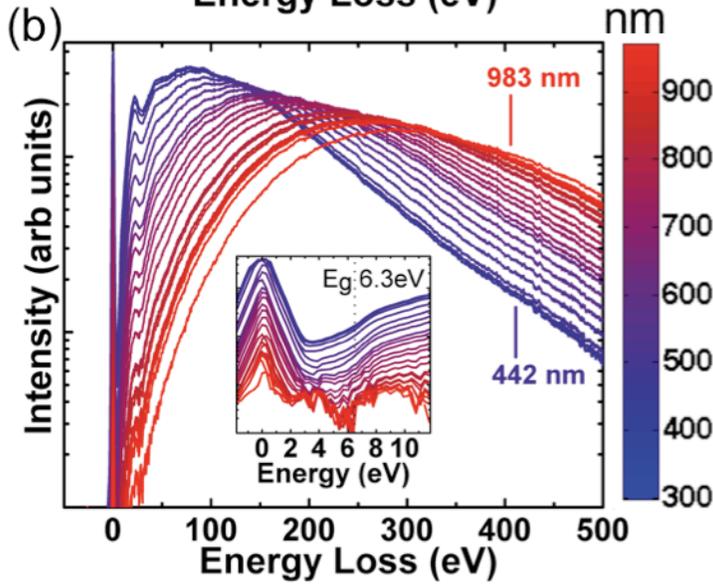

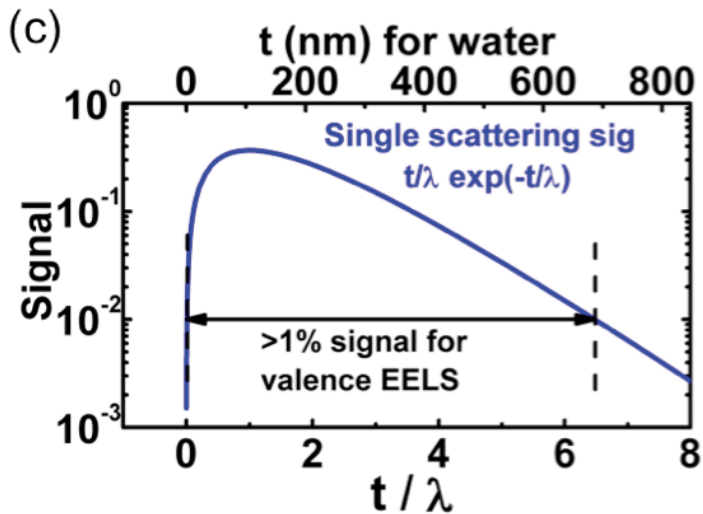

**Figure 2.** STEM-EELS measurements through (a) water as a function of the thickness of the liquid: the O-K edge is resolvable for the thinnest layer only while the optical gap at 6.9 eV is resolvable up to ~650 nm (6.5 $\lambda$). Each spectrum is a single acquisition, so the core-loss and low-loss are acquired simultaneously, and the thickness is determined from the complete spectrum. (b) EELS for a wide range of thicknesses of ethylene glycol, showing the optical gap at 6.3 eV is resolved up to ~600 nm (6 $\lambda$). The color of the spectra corresponds to the thickness of the liquid as denoted on the color bar in (a-b). (c) The first order Poisson distribution shows that >1% of the signal is singly scattered for thicknesses less than $t/\lambda = 6.5$, giving an estimate of the feasible range of operation for valence EELS, which agrees with the insets of (a-b).

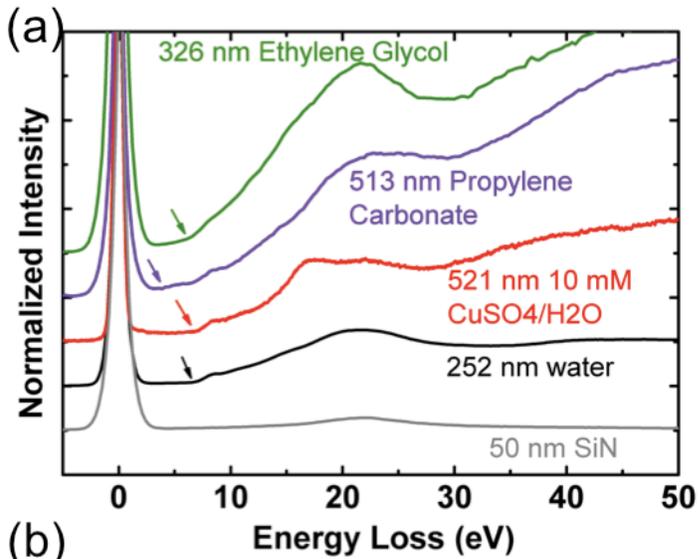
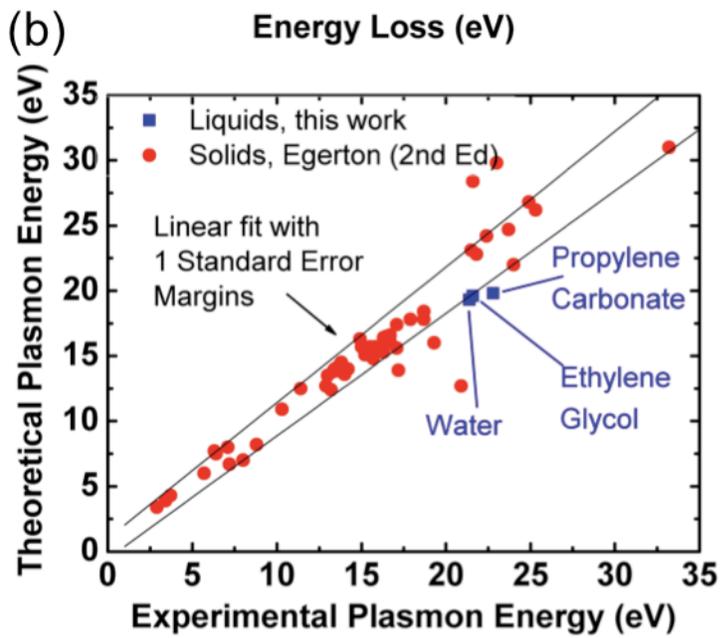

**Figure 3.** (a) Valence EELS of ethylene glycol, propylene carbonate and 10 mM of $CuSO_4/H_2O$. Arrows point to the optical gap. The background spectrum from the $SiN_x$ membranes alone shows no dominant features on the same scale without the presence of liquid. (b) Theoretical free-electron estimates of the plasmon energy compared to experimental plasmon energy for solids (Egerton, 1986) and our measurements of the pure liquids. The solid lines are the one-sigma standard error margins of a linear fit.

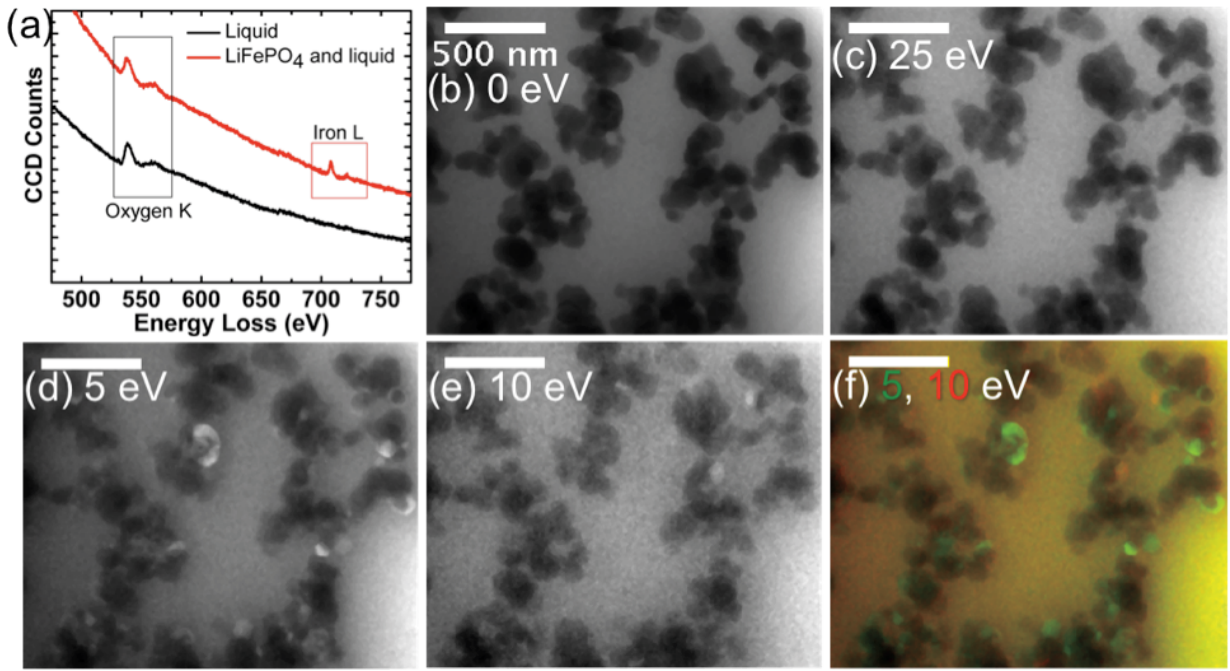
25

**Figure 4.** STEM-EELS and EFTEM of LiFePO$_4$ nanoparticles 180 nm of aqueous solution. (a) Core-loss STEM-EELS on and off a nanoparticle, where the Fe L-edge is seen on the particle and the O K-edge is seen both on the particle and in the solution. EFTEM with a 5 eV energy selecting slit centered around (b) 0 eV where the liquid dominates the signal, (c) 25 eV at the plasmon peak of the liquid, where the liquid dominates the signal, (d) 5 eV in the optical gap of the liquid showing FePO$_4$ transitions, (e) 10 eV, with a low background signal from the liquid and a slightly stronger signal from LiFePO$_4$ than FePO$_4$, and (f) an overlay of the 5 eV image in green highlighting the FePO$_4$ and the 10 eV image in red showing the LiFePO$_4$.

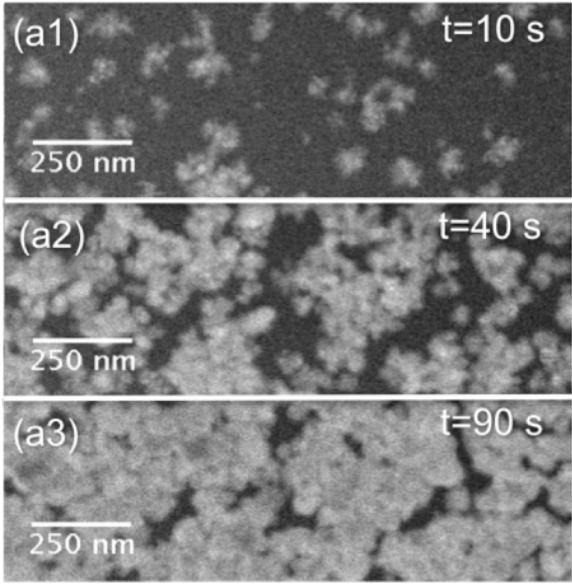
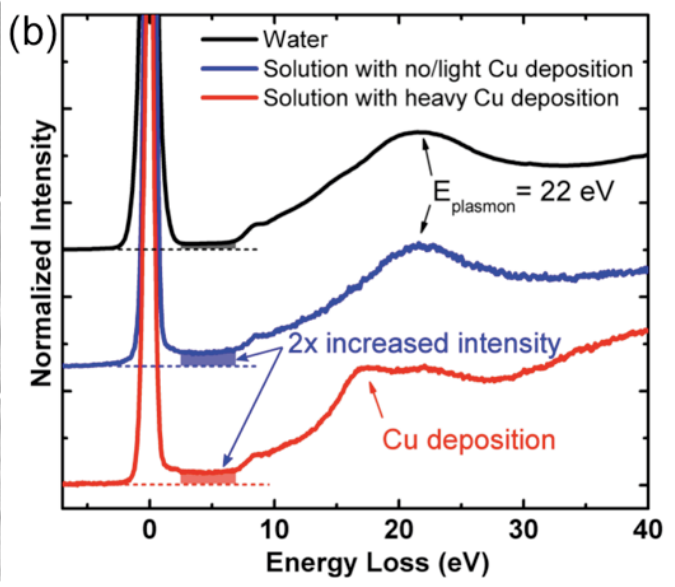

**Figure 5.** *In situ* STEM imaging of beam-induced Cu deposition in 10 mM $CuSO_4$/$H_2O$ after (a1) 10 s, (a2) 40 s and (a3) 90 s, at a 30 e-/nm$^2$·s dose rate. (b) STEM-EELS after light Cu deposition (~2s) and heavy Cu deposition (~90s) are compared to water. The 10 mM $CuSO_4$ solution shows increased Cherenkov radiation compared to that of $H_2O$. After prolonged deposition, the EELS spectrum develops a peak at 17 eV, which is characteristic of bulk Cu.

TABLES.

**Table 1.**

Electronic structure properties found from valence EELS of four different liquids. The optical gap and the plasmon energy are found from Fig. 2. The local electron density and the inelastic mean free path are calculated from the plasmon energy and Eq. 3 respectively. The optical gap determined by UV VIS was determined and shown in the supplementary information (Fig. S3), or noted from the CRC Handbook.

| Liquid | Theoretical Plasmon Energy (eV) | Plasmon Energy by EELS (eV) | Measured Local Electron Density ($cm^{-3}$) | Inelastic Mean Free Path (nm) (Eq.2) | Optical Gap by EELS (eV) | Optical Gap by UV-VIS (eV) |
|---|---|---|---|---|---|---|
| Water | 19.3 | 21.4 | $3.30 \times 10^{23}$ | 106 | 6.9 | 6.5* |
| 10mM CuSO4/Water | -- | 22.1 | $3.52 \times 10^{23}$ | 103 | 5.8 | 4.1 |
| Propylene Carbonate | 19.8 | 22.8 | $3.75 \times 10^{23}$ | 100 | 4.2 | 5.2 |
| Ethylene Glycol | 19.6 | 21.6 | $3.36 \times 10^{23}$ | 105 | 6.3 | 6.3 |

- *from CRC Handbook of Chemistry and Physics(Haynes & Lide, 2011)

ACKNOWLEDGEMENTS

We would like to thank John Grazul and Mick Thomas for assistance in the TEM facility. This work made use of the electron microscopy facility of the Cornell Center for Materials Research (CCMR) with support from the National Science Foundation Materials Research Science and Engineering Centers (MRSEC) program (DMR 1120296). This work and MEH and YY are supported by the Energy Materials Center at Cornell, an Energy Frontier Research Center funded by the U.S. Department of Energy, Office of Basic Energy Sciences under Award Number DESC0001086.